# Coherent long-distance displacement of individual electron spins

H. Flentje[1,2], P-A. Mortemousque[1,2], R. Thalineau[1,2], A. Ludwig[3], A. D. Wieck[3], C. Bäuerle[1,2], T. Meunier[1,2] *

[1] *Univ. Grenoble Alpes, Institut NEEL, F-38042 Grenoble, France.*

[2] *CNRS, Institut NEEL, F-38042 Grenoble, France.*

[3] *Lehrstuhl für Angewandte Festkörperphysik, Ruhr-Universität Bochum, Universitätsstraße 150, D-44780 Bochum, Germany.*

**Controlling nanocircuits at the single electron spin level is a possible route for large-scale quantum information processing[1,2]. In this context, individual electron spins have been identified as versatile quantum information carriers to interconnect different nodes of a spin-based semiconductor quantum circuit[2]. Despite important experimental efforts to control the electron displacement over long distances[3,4,5], keeping the electron spin coherence after transfer remained up to now elusive. Here we demonstrate that individual electron spins can be displaced coherently over a distance of 5 μm. This displacement is realized on a closed path made of three tunnel-coupled lateral quantum dots. Using fast quantum dot control, the electrons tunnel from one dot to another at a speed approaching 100 m/s. We find that the spin coherence length is 8 times longer than expected from the electron spin coherence without displacement. Such an enhanced spin coherence points at a process similar to motional narrowing observed in nuclear magnetic resonance**

**experiments[6]. The demonstrated coherent displacement will enable long-range interaction between distant spin-qubits and will open the route towards non-abelian and holonomic manipulation of a single electron spin[7, 8].**

While it is clear by now that the spin degree of freedom of an electron is an interesting building block for processing and storing quantum information[9-11], important questions concerning the system scalability remain to be addressed before building a large scale spin-based quantum processor. Ultimately, the problem reduces to the ability to transfer quantum information on a chip. Following the work on superconducting qubits, a significant experimental effort is currently focusing on the possibility to couple distant electron spins via a quantum mediator[12-14]. An alternative way consists in displacing the electron spin itself[2]. One possibility is to convey the electron in moving quantum dots defined by surface acoustic waves, where it is trapped and propagates isolated from the surrounding electrons at the speed of sound[5,15-17]. Even though electron and spin transfer have been demonstrated, the technology of moving quantum dots at the single electron level is not yet controlled well enough to investigate coherence properties[5]. A more conventional strategy consists in displacing the electron in an array of tunnel-coupled quantum dots[3,4,18-23]. So far, only classical spin transfer over linear arrays of three and four dots has been demonstrated[3,4,23], whereas slow electron displacement on a closed loop has been demonstrated in a four-quantum-dot system[19].

To demonstrate the coherent spin transfer of individual electrons, we have investigated the spin dynamics of two electrons initially prepared in a singlet spin state and displaced in an array of three lateral quantum dots defined in a circular geometry within an AlGaAs heterostructure (see Fig. 1a). The dot system is tuned in the isolated configuration where the coupling to the electron reservoir can be ignored (see Supplementary Section I)[24]. The two electrons are loaded into the system via the bottom dot. The resulting charge response of the electrometer when changing the chemical potentials of the three dots is presented in Fig. 1b. As expected, only possible six charge configurations are observed (see Supplementary Section I), rendering electron displacement on a closed loop trivial. The tunnelling rates between the three dots are tunable up to the gigahertz regime. Nanosecond control of the gate voltages permits therefore adiabatic electron transfer between the dots faster than the spin coherence time[24].

The two-electron spin state after displacement can be inferred by bringing the electrons in the bottom dot where exchange of electrons with the reservoir is possible[24]. In the two-electron case, the ground singlet (S) and the three excited triplet ($T_+$, $T_0$, $T_-$) states are distinguished using the tunnel-rate spin read-out method with a single-shot fidelity of 80% (see Supplementary Section II)[25,26].

First, we specifically focus on the spin dynamics when the two electrons are static in two different dots. Initialization in the singlet ground state is performed by relaxation in the bottom dot. By rapidly pulsing the gate voltages, the electrons are separated into two different dots for a controlled duration $\tau_s$. In this way, we probe how long the phase coherence initially present in the singlet state can be preserved when the electrons are separated[18]. If the phase coherence is preserved, the system will remain in the singlet state. Otherwise the final spin state will be a mixture of singlet and triplet states. Figure 2a presents the spin singlet population when the charge stability diagram is explored with a 50-ns voltage pulse on $V_1$ and $V_2$. Three distinct regions where the spin mixing is efficient are observed. They correspond to the three charge configurations where the electrons are separated in two dots. In these regions, the exchange interaction between the two electrons can be neglected and the system is dominated by the coupling to the nuclear spins of the heterostructure via hyperfine interaction[18,24,27]. At a magnetic field of 150 mT, the spin mixing occurs only between S and $T_0$ (see Fig. 2d). By varying $\tau_s$, we observe a Gaussian decay of the singlet probability with typical timescales close to 10 ns, very similar in each mixing region[18,24]. In addition to the S-$T_0$ mixing regions, we notice four additional mixing lines that are expected from the mixing of S and $T_+$ states when the tunnel-coupling between the dots is large and coherent (see Supplementary Section III)[24].

We proceed to the investigation of the two-electron phase coherence while the two electrons are individually displaced on the closed loop formed by the three quantum dots. More specifically, the electrons are initially prepared in the singlet state of the (2, 0, 0) charge configuration. The system is then pulsed fast to the region (1, 1, 0) where the electrons are separated in two dots and rotated repeatedly between the spin mixing regions of the (1, 1, 0), (0, 1, 1) and (1, 0, 1) charge configurations with voltage pulse sequence on $V_1$ and $V_2$ (see Fig. 2a and 2b). It leads to a series of quantum dot displacements and single electron tunnelling events schematically shown in Fig. 2c. Arbitrary long displacements can be implemented by repeating the loop. We control the number of loops $N_t$ performed by the electrons and the duration $\tau_r$ spent in each charge configuration. Finally, the system is tuned back from the (1, 1, 0) to the (2, 0, 0) charge configuration where spin read-out is performed. With $\tau_r$ equal to 1.7 ns, the resulting singlet probability is exponentially decaying as a function of $\tau_s = 3N_t\tau_r$, the time spent in configurations where the electrons are separated. These results demonstrate coherent electron spin transfer in an array of quantum dots in a circular geometry. We measured a spin coherence time of 80 ns, almost 8-times longer than for the static case (see Fig. 2e), only possible with a significant reduction of the influence of the hyperfine interaction during the electron displacement.

For $\tau_r$ set to 2.5 ns, the time dependence of the spin mixing no longer exhibits a single exponential behaviour but is characterized by two timescales (see Inset Fig. 2e). First, the system decays fast on a timescale similar to the coherence time in static dots, and then it evolves on a longer timescale towards a mixed singlet-triplet state. Figure 3a shows the spin singlet probability as a function of $N_t$ for different values of $\tau_r$. The long decay is only dependent on $N_t$. We interpret these observations as the consequence of two different phases during the displacement procedure: the "static" phase where the electrons are static in two different dots and the "transfer" phase where they are moving between two dots separated by approximately 110 nm on a fixed timescale corresponding to the rise time of the pulse generator (0.9 ns).

During the "static" phase, the electrons are experiencing a fast spin mixing induced by hyperfine interaction. When the electrons realize only one rotation with increasing $\tau_r$, the influence of the transfer phase is minimized. In Fig. 3b, the observed spin mixing time is 1.74 ± 0.17 times longer than in the static configuration. This increase is very close to the $\sqrt{3}$-factor expected for an electron spin coupled to a 3-times larger nuclear spin bath via hyperfine interaction[27,28].

During the "transfer" phase, the electrons are displaced in moving quantum dots induced by the time-dependent potentials applied on the gates, before and after the tunnelling processes. As a consequence, the number of nuclei coupled to the electrons is drastically increased. Considering an electron displacement velocity close to 100 m/s, the hyperfine-limited coherence time is expected to increase up to the µs-timescale in a process similar to motional narrowing observed in liquid nuclear magnetic resonance experiments[6,28]. Spin-flip processes, stimulated by the electron motion and resulting from either spin-orbit or transverse hyperfine couplings, are then expected to limit the spin coherence time[28].

For $\tau_r$ set to 1.7 ns, the time spent in the static phase is minimized and the spin decoherence process is mainly occurring during the transfer phase (see Methods). In this case, the spin dynamics is characterized by a single exponential decay of the singlet probability (see Fig. 4). As the magnetic field is increased from 0 to 200 mT, the individual spin-flip processes are expected to become less and less efficient. It results in a progressive reduction of the singlet mixing with $T_+$ and $T_-$ after a 250-ns evolution (see Supplementary Section IV). Furthermore, we observe a linear increase of the spin coherence time, from 12 ns to almost 80 ns. Considering the estimated distance of 110 nm between the dots (see Supplementary Section I), we measure a maximal spin coherence length of 5 µm at 200 mT.

In conclusion, we have demonstrated that the coherence of a two-electron singlet state is preserved when the electrons are separated and displaced over 5 µm on a closed loop in a three-dot system. Compared to the situation without displacement, the spin coherence time is increased by a factor of 8 via a motional narrowing process and is equal to 80 ns. Furthermore, spin-flip processes stimulated by the electron motion are found to limit the spin coherence time. The demonstrated coherent spin displacement could be a viable route to interconnect quantum nodes in spin-based quantum processors. On a more fundamental side, increasing the speed of the closed-loop transfer with larger tunnel-couplings should allow to explore non abelian and holonomic spin manipulation[7, 8, 29, 30] in future experiments.

**METHODS:**

The device is defined by Schottky gates in an n-$Al_{0.3}Ga_{0.7}As$/GaAs 2DEG-based heterostructure (the properties of the non-illuminated 2DEG are as follows: mobility $\mu \approx 10^6$ cm$^2$ V$^{-1}$ s$^{-1}$, density $n_s \approx 2.7 \times 10^{11}$ cm$^{-2}$, depth 100 nm) with standard split-gate techniques. It is anchored to a cold finger mechanically screwed to the mixing chamber of a dilution fridge with a base temperature of 70 mK. It is placed at the center of the magnetic field produced by a solenoid. The coil allows to produce magnetic fields perpendicular to the 2DEG. The charge configuration of the triple-dot system is determined by measuring the conductance of the sensing dots biased with 300 µV; the current is measured using a current-to-voltage converter with a bandwidth of 10 kHz. The voltage on each gate can be varied on µs-timescales to allow exploration of the isolated configuration. Each green gate ($V_1$, $V_2$, $V_3$) in Fig. 1a is connected through a low temperature home-made bias-T to both DC and high bandwidth coaxial lines allowing gigahertz manipulations. The voltage pulses to induce electron displacement are generated by an arbitrary waveform generator Tektronix 5014C with a typical rise time (20% - 80%) approaching 0.9 ns. For $\tau_r$ equal to 1.7 ns, the pulse sequence presented in Fig. 2b is just reaching the programmed voltage amplitude. We can therefore assume that the electrons are only in the "transfer" phase during the displacement for $\tau_r = 1.7$ ns.

**Acknowledgments:**

We acknowledge technical support from the "Poles" of the Institut Néel as well as from Pierre Perrier and Henri Rodenas. A.L. and A.D.W. acknowledges gratefully the support of the BMBF Q.com-H 16KIS0109, Mercur Pr-2013-0001 and the DFH/UFA CDFA-05-06. T.M. acknowledges financial support from ERC "QSPINMOTION".

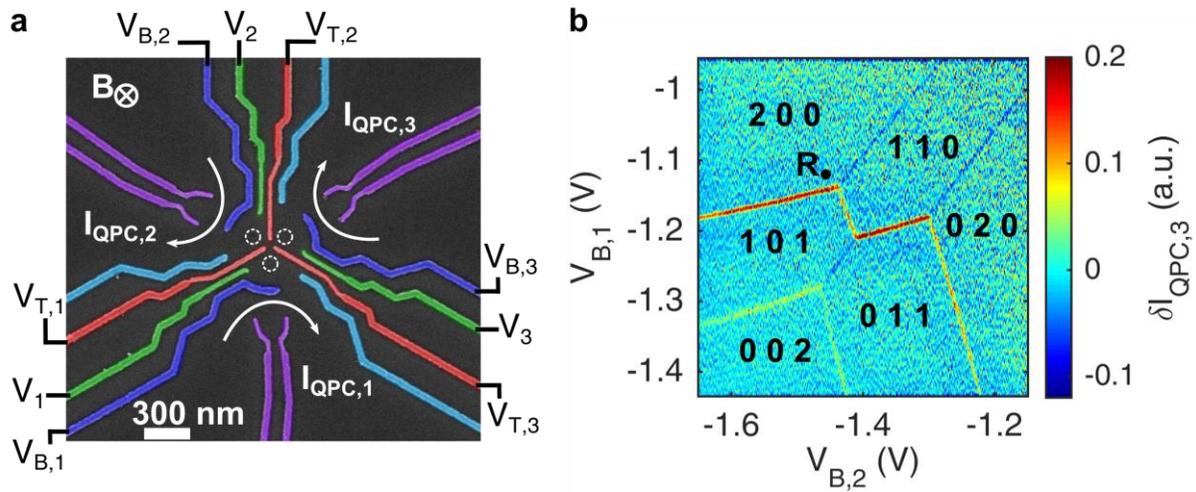

**Figure 1 Experimental set-up and two-electron stability diagram in the isolated configuration. a,** Scanning electron microscope image of the circular triple dot sample. The position of the dots are shown in white dashed circles. The device is defined by Schottky gates in an AlGaAs 2DEG-based heterostructure with standard split-gate techniques (see Methods). The voltages applied on the green ($V_1,V_2,V_3$), red ($V_{T,1},V_{T,2},V_{T,3}$) and blue gates ($V_{B,1},V_{B,2},V_{B,3}$) allow to predominantly control the coupling between the dots, the coupling to the reservoirs and the dot-chemical potentials respectively. The purple gates are used to define sensing dots to probe, with $I_{QPC,1}$, $I_{QPC,2}$ and $I_{QPC,3}$, the charge configuration of the triple-dot system. Electron loading and spin read-out are realized in the bottom dot. A magnetic field B is applied perpendicular to the sample. **b,** Derivative $\delta I_{QPC,3}$ of $I_{QPC,3}$ along $V_{B,1}$ when the system is scanned in the two-electron isolated configuration with the gates $V_{B,1}$ and $V_{B,2}$. The label ($N_1$, $N_2$, $N_3$) corresponds to the number of electrons in the bottom, top left and top right dots respectively.

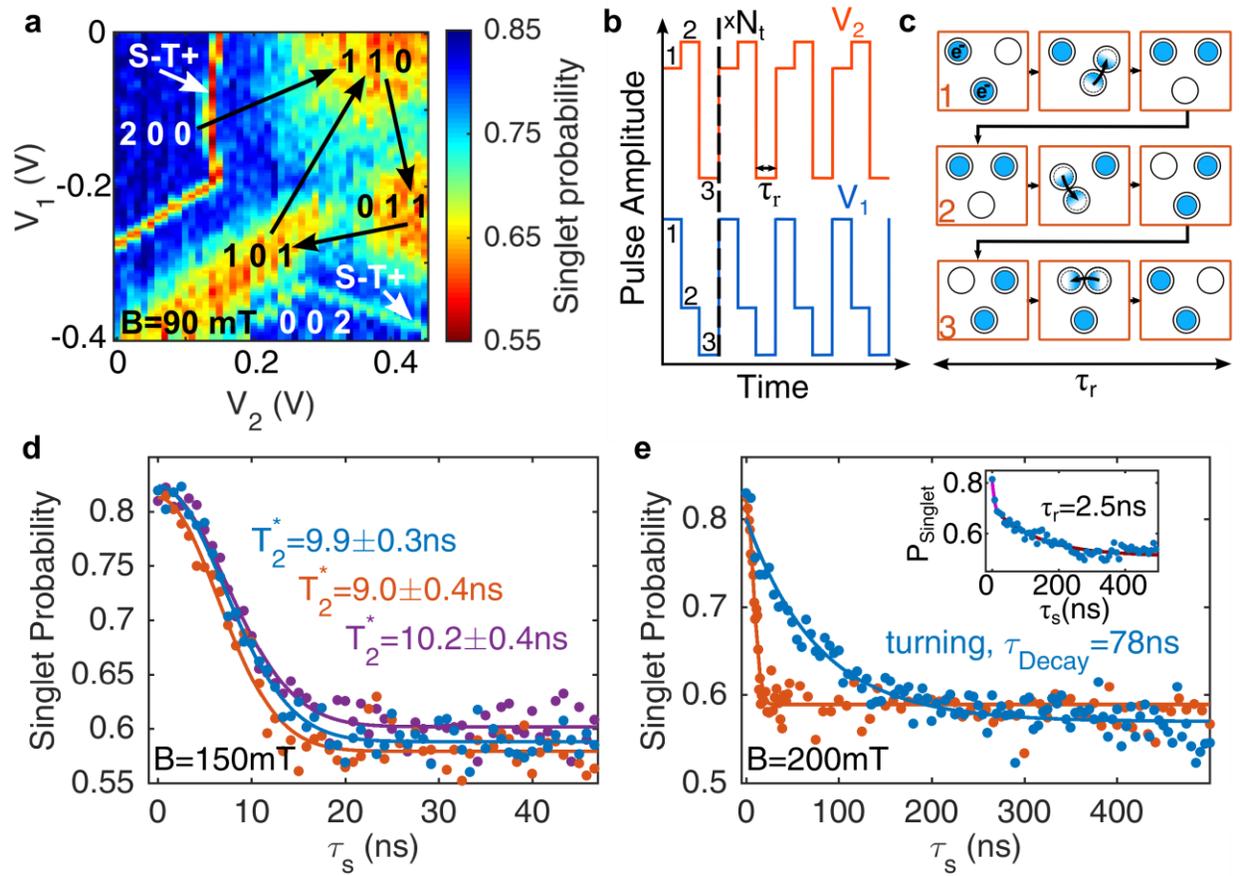

**Figure 2 Coherent spin displacement. a,** Two-electron spin mixing map. From the position R in Fig. 1b, a 50-ns pulse on $V_1$ and $V_2$ is applied before performing single-shot read-out of the two-electron spin states to extract the singlet probability. In the separated configurations, S mixes with $T_0$. In between two of these regions, one electron is exchanged between two dots and the spin mixing is less effective due to the increase of the exchange interaction (see Supplementary Section III). The arrows represent the typical path taken by the electron during the coherent spin transport. **b,** Schematics of the time-dependent sequence applied on gates $V_1$ and $V_2$ to perform the electron displacement. $N_t$ is the number of loops performed by the electrons and $\tau_r$ is the duration spent in each charge configuration (1 = (1, 1, 0), 2 = (0, 1, 1), 3 = (1, 0, 1)).

Considering the rise time of the pulse generator, the electrons are adiabatically transferred between the dots (see Methods). **c,** Schematics of the spatial displacement of the electrons during a single rotation. Most of the displacement is occurring while the electrons are trapped in moving quantum dots. **d,** Singlet probability as a function of the time $\tau_s$ spent in separated configurations where the electrons are static in (0, 1, 1), (1, 1, 0) and (1, 0, 1), (orange, blue and purple respectively). The data are fitted with a Gaussian decay with a characteristic time $T_2^*$. **e,** Singlet probability as a function of the time $\tau_s$ for the case where the electrons are rotating between separated charge configurations with $\tau_r$ = 1.7 ns (blue). The data are fitted with an exponential decay with a characteristic time $\tau_{Decay}$. The curve without displacement in the (0, 1, 1) charge configuration is reproduced in orange for comparison. Inset: Singlet probability as a function of the time $\tau_s$ for B = 90 mT and $\tau_r$ = 2.5 ns. The data are fitted with a two exponential decays.

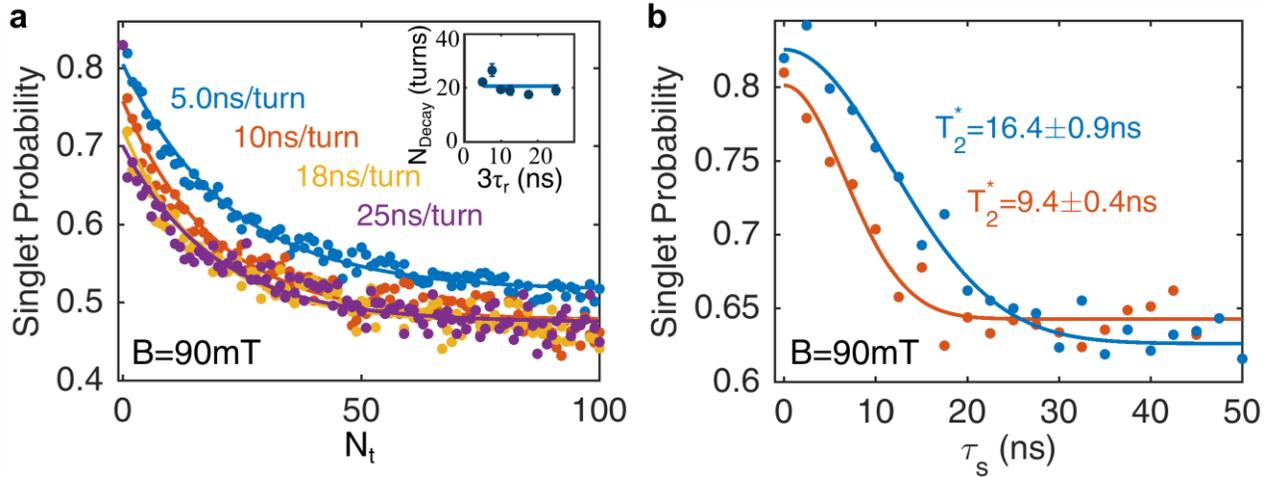

**Figure 3 Influence of the number of turns on spin mixing and motional narrowing**

**a,** Singlet probability as a function of the number of turns $N_t$ for different time per turn $3\tau_r$. The data after the static phase are fitted with an exponential decay with a characteristic number of turns $N_{Decay}$. Inset: Extracted $N_{Decay}$ as a function of $3\tau_r$. The solid line is a constant fit to the data which slightly depends on the voltage gate configuration (see Supplementary Section V). **b,** Singlet probability as a function $\tau_s$ for $N_t$ equal to one and obtained by increasing $\tau_r$. The corresponding singlet probability as a function of $\tau_s$ where the electrons are static in the (0, 1, 1) charge configuration is reproduced in orange for comparison. The data are fitted with a Gaussian decay with a characteristic time $T_2^*$.

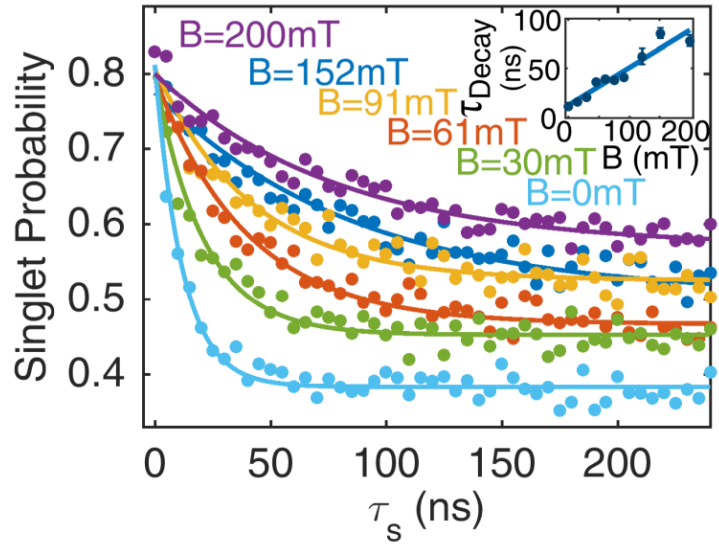

**Figure 4 Influence of the magnetic field on the spin coherence time.** Singlet probability as a function of the time $\tau_s$ for the case where the electrons are rotating in the triple-dot system for different B at fixed $\tau_r$ = 1.7 ns. The data are fitted with an exponential decay with a characteristic time $\tau_{Decay}$. Inset: Extracted $\tau_{Decay}$ as a function of the magnetic field B. The solid line is a linear fit of the data.

**Supplementary information for**

# Coherent displacement of individual electron spins over 5 µm

H. Flentje, P-A. Mortemousque, R. Thalineau, A. Ludwig, A. D. Wieck, C. Bäuerle, T. Meunier

# Content

# I. Circular triple-dot in the isolated configuration

In the main text, the manipulation and the control of two electrons in the isolated configuration of a circular triple-dot system is described. In this section, we provide an electrostatic simulation of the potential experienced by a single electron (see Fig. S1a), and we describe in details how to reach the isolated configuration and the stability diagram corresponding to different numbers of electrons.

In the isolated configuration, we are able to characterize the triple-dot system with a fixed number of electrons over a wide range of gate voltages. Following the procedure presented in ref 24, it is indeed possible to load first the bottom dot with the desired number of electrons and then rapidly promote them into the isolated position with a microsecond pulse mainly applied on the blue gates of Fig. 1a (see Fig. S1b). Finally, the system is scanned from that position to reconstruct a stability diagram of the isolated double dot. Figure S1c (1b) shows the observed stability diagram with the overall electron number fixed to one (two). They are characterized by three (six) distinct charge configurations separated by inter-dot degeneracy lines. These different charge configurations are the only possible charge states with a fixed number of electrons loaded in the isolated configuration[24].

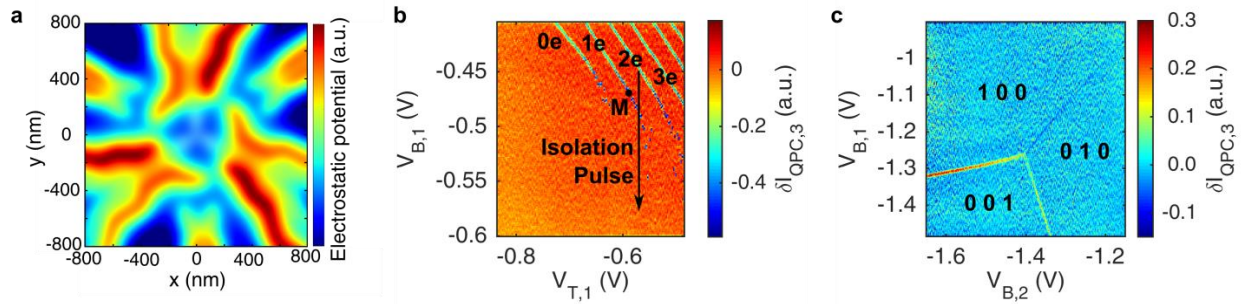

**Supplementary Fig. S1. Circular triple quantum dot in the isolated regime. a,** Electrostatic potential experienced by the electron in a working configuration showing the three quantum dots in a circular configuration. The calculation follows the procedure presented in ref 31. The distance between two dots is estimated to be 110 nm. **b**, Stability diagram of the bottom quantum dot close to the isolated regime. Derivative $\delta I_{QPC,3}$ of $I_{QPC,3}$ when the system is scanned with the gates $V_{B,1}$ and $V_{T,1}$. The charge degeneracy lines are vanishing because the tunnelling time to the reservoir is becoming progressively longer as $V_{B,1}$ becomes more negative. The loading positions for one, two, three electrons and the spin measurement position are indicated with 1e, 2e, 3e and M, respectively. **c**, Isolated stability diagram for the case of one electron initialized in the system. Derivative $\delta I_{QPC,3}$ of $I_{QPC,3}$ when the system is scanned in the one-electron isolated configuration with the gates $V_{B,1}$ and $V_{B,2}$. The dot occupation numbers are given in the graph following the notation of Fig. 1b.

## II. Tunnel-rate spin read-out in the isolated configuration and its fidelity

To perform spin read-out of the two-electron spin states in the isolated configuration, the two electrons are brought back into the bottom dot in a configuration where the electrons can be exchanged with the reservoir[24]. In this section, a detailed description of the spin read-out procedure and an evaluation of the spin read-out fidelity is given.

To probe the spin dynamics of the electrons in the triple-dot system, a spin readout protocol compatible with the electron displacement procedure has been implemented. The principle to detect electron spin states in a single quantum dot coupled to a lead is well established[9,25,26]. It relies on the engineering of a spin-dependent tunnel process from the dot to the reservoir to convert spin into charge information. To perform the single shot readout of a two-electron spin state, we take advantage of the difference in tunnel-rates for singlet ($\Gamma_S$) and triplet states ($\Gamma_T$) at the measurement position M in Fig. S1b where one electron is allowed to tunnel out of the dot (see Fig. S2a and b). We estimate the ratio $\Gamma_T/\Gamma_S$ to be 10 in our experiment fixed by the shape of the dot. However, the scheme used to displace the electrons requires working with the triple-dot system in an isolated configuration, where no exchange of electrons is possible between the dot and the reservoir[24]. For this reason, we have to bring back the dot system to the measurement position at a µs-timescale (much faster than the spin relaxation time) to infer the electron spin state in the isolated configuration.

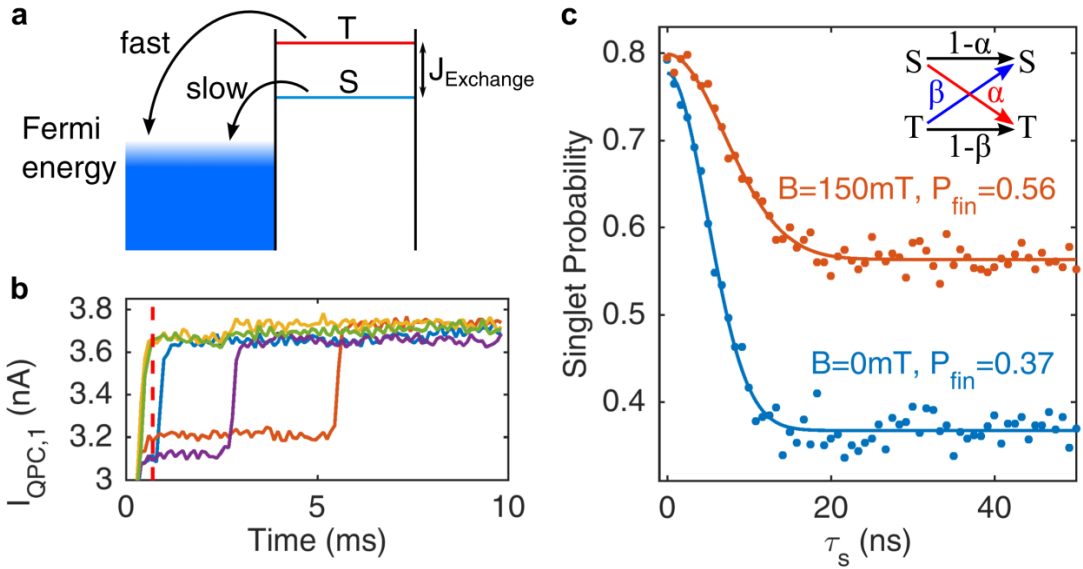

**Supplementary Fig. S2 Tunnel-rate dependent spin readout. a,** Principle of the spin read-out based on the difference of the tunnel-rate to the reservoir between the singlet S and the triplet T states. **b,** Typical single-shot current time-traces used to distinguish S and T states. An electron tunnelling out of the dot results in a current jump from the two-to-one electron charge configuration. The red dashed line indicates the time-threshold condition for triplet identification. **c,** Singlet probability as a function of $\tau_S$ measured in the isolated position where hyperfine interaction is dominant at zero (blue) and 150 mT (red) external magnetic field. A Gaussian fit (solid lines) reveals a mixing time equal to 6 ns and 10 ns respectively. The final spin mixing values $P_{fin}$ allow us to estimate the errors $\alpha$ and $\beta$ in the detection of the singlet and the triplet states according to the error scheme shown in the inset.

To quantify the fidelity of the complete spin read-out procedure, we have analysed the evolution of the singlet probability as a function of the time $\tau_S$ where the electrons are separated at different magnetic fields (see Fig. S2c). At zero magnetic field, all the three triplet states should mix with the singlet state whereas only $T_0$ mixes with the singlet state at 100 mT. The data are fitted with a Gaussian decay. The observed loss of coherence is explained by the fluctuating effective magnetic field difference between the dots of 4 mT due to the hyperfine interaction. The initial and steady-state probability values of the decay curves allow extracting the measurement errors $\alpha$ and $\beta$ for singlet and triplet respectively. The electron should start in singlet states for both magnetic field conditions whereas the proportion of singlet at the end of the mixing should be 0.5 at 0.15 T and 0.3 at 0 T[27]. Such an analysis gives us an average measurement fidelity of 1 - $(\alpha+\beta)/2$ = 80%.

## III. Model of the triple dot system with two electrons

A spin map procedure has been implemented to identify where spin-mixing between the singlet and the triplet states occurs in the gate voltage space. Such a mixing is the result of a competition between the hyperfine and the exchange interactions. In a double dot[9], the level repulsion induced by tunnelling only affects the singlet states and results in an energy splitting $J_{exchange}$, the exchange energy (see Fig. S2a), between singlet and triplet states. In this section, the three quantum dot system, circularly tunnel-coupled, with two electrons is modelled in order to simulate the result of the spin mixing map procedure.

In this simulation, the hyperfine interaction is modelled as an effective magnetic field $B_n$ in each dot that is fluctuating between two realizations of the experiment with a standard deviation $\Delta B_n$= 2.8 mT. The relevant parameters of the Hamiltonian are then the effective magnetic field in each dot, the tunnel-couplings between the dots assumed to be equal, and the chemical potential and the charging energy of each dot. In each gate voltage configuration, we plot the result of the 50-ns spin evolution starting from the singlet state with an averaging on the fluctuating effective magnetic field conditions. The results of the simulation is presented in Fig S3. It reproduces qualitatively the main features of the measured spin map. In particular, an increase of the singlet population in the region where the two electrons are separated is observed when the system is close to the degeneracy between two dots, for example the (1,1,0)-(1,0,1) crossing region. In a circularly coupled triple-dot, up to three tunnel-couplings are indeed contributing to the repulsion of the singlet states and an increase of the exchange interaction close to the separated degeneracy line is expected.

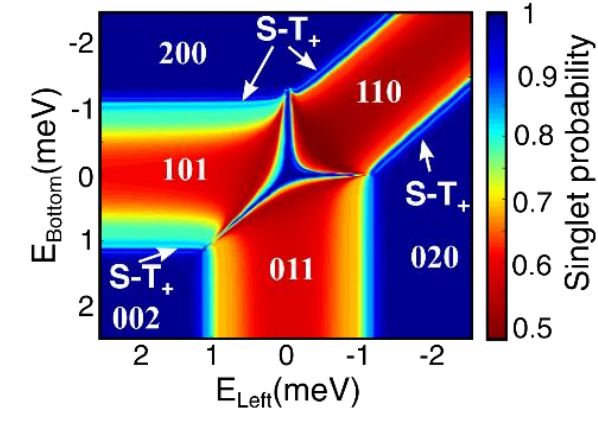

**Supplementary Fig. S3 Simulated spin mixing map for the triple quantum dot with two electrons.** Singlet probability after 50 ns of free evolution as a function of the energy of the left ($E_{Left}$) and bottom quantum dot ($E_{Bottom}$). The energy of the right quantum dot was chosen to be $E_{Right} = 0$. The charge configuration is first inferred from the lowest energy eigenstates of the Hamiltonian and the appropriate singlet state is selected. Then a time evolution for the respective energy configuration is performed and the singlet probability is averaged over Gaussian distributed nuclear field values in each quantum dot. In this diagram, the tunnel-coupling energy is $t = 12$ µeV and the Zeeman energy splitting from the magnetic field is $E_Z = 4$ µeV.

# IV. Increase of the spin coherence time while the two electrons are moving at zero magnetic field

The increase in spin coherence time at a magnetic field of 200 mT is precisely analysed in the main text and a strong dependence of the spin coherence with magnetic field is shown.

In this section, we present the investigation of the impact of displacement on the spin mixing at zero magnetic field. We have implemented the same procedure as discussed in Fig. 2 at zero magnetic field. The resulting singlet probabilities as a function of the time $\tau_S$ where the electrons are separated are presented in Fig. S4. Without displacing the electrons, the singlet probability is characterized by a Gaussian decay with a typical timescale close to 6 ns. Such a reduction of the spin coherence time at zero magnetic field is expected since all the three triplet states can be mixed with the singlet state due to hyperfine interaction and the three components of the effective magnetic field have to be taken into account[27]. With the electron displacement, an exponential decay of the singlet probability is observed with an increase of the spin coherence time by a factor of two.

We conclude that the spin-flip process is more efficient at zero magnetic field. In this situation, the electrons do not need an energy exchange with a reservoir to flip their spin. In the manuscript, two mechanisms were identified as the main sources of decoherence during electron displacement: spin-orbit and transverse hyperfine interactions. At zero field, the spin-orbit interaction results in a coherent evolution of the electron spin during its displacement on a fixed path[7,8]. The singlet state is then expected to be preserved along the displacement. To be in agreement with the data, changes of the electron path due to the large microwave excitation could result in a fast mixing of the singlet state with the triplet states. It is worth noting that the transverse hyperfine interaction is in essence a fluctuating coupling between the spin and the motion and, on the contrary to the spin-orbit interaction, path fluctuations are therefore not required to explain the fast mixing of the singlet state with triplet states.

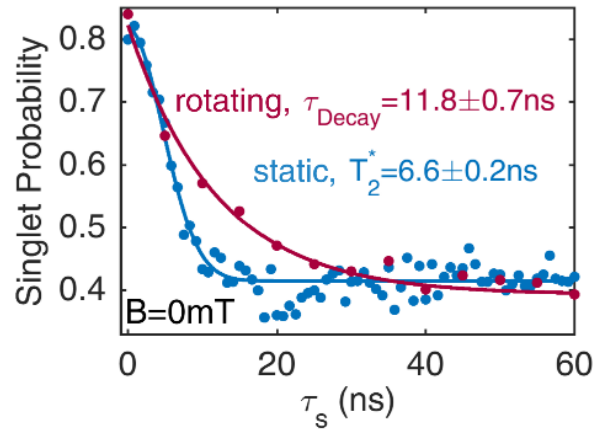

**Supplementary Fig. S4 Coherent spin displacement at zero magnetic field.** Singlet probability as a function of the time $\tau_s$ for the case where the electrons are static in the (0,1,1) charge configuration (blue) and where they are rotating between separated charge configurations with $\tau_r$ = 1.7 ns (red). The data are fitted with a Gaussian decay with a characteristic time $T_2^*$ for the static procedure. The case for rotational movement shows better agreement with an exponential decay with a characteristic time $\tau_{Decay}$ and shows a significantly longer decay time constant.

# V. Influence of the tunnel-coupling on the displacement-induced spin coherence time

In the main text, the presented results have been mostly obtained for a specific tunnel-coupling between the dots. In this section, the spin coherence time after displacement for different tunnel-couplings between the dots is analysed. The strength of the tunnel-couplings can be changed by controlling the potential of the red gates in Fig. 1a. It can directly be witnessed on the spin mixing maps in Fig S5b-d: for decreasing tunnel-coupling, the separation between the three mixing regions is progressively vanishing. We observe that the exchange interaction at the single-electron crossing between two dots when the electrons are separated is reduced and becomes negligible in comparison with the hyperfine interaction. Moreover, the separation in gate space between the S-T$_0$ mixing region and the S-T+ crossing lines is progressively reduced until it vanishes completely. Such observations are in agreement with a progressive reduction of the singlet level repulsion induced by the tunnelling process and are consistent with a reduction of the tunnel-couplings between the dots.

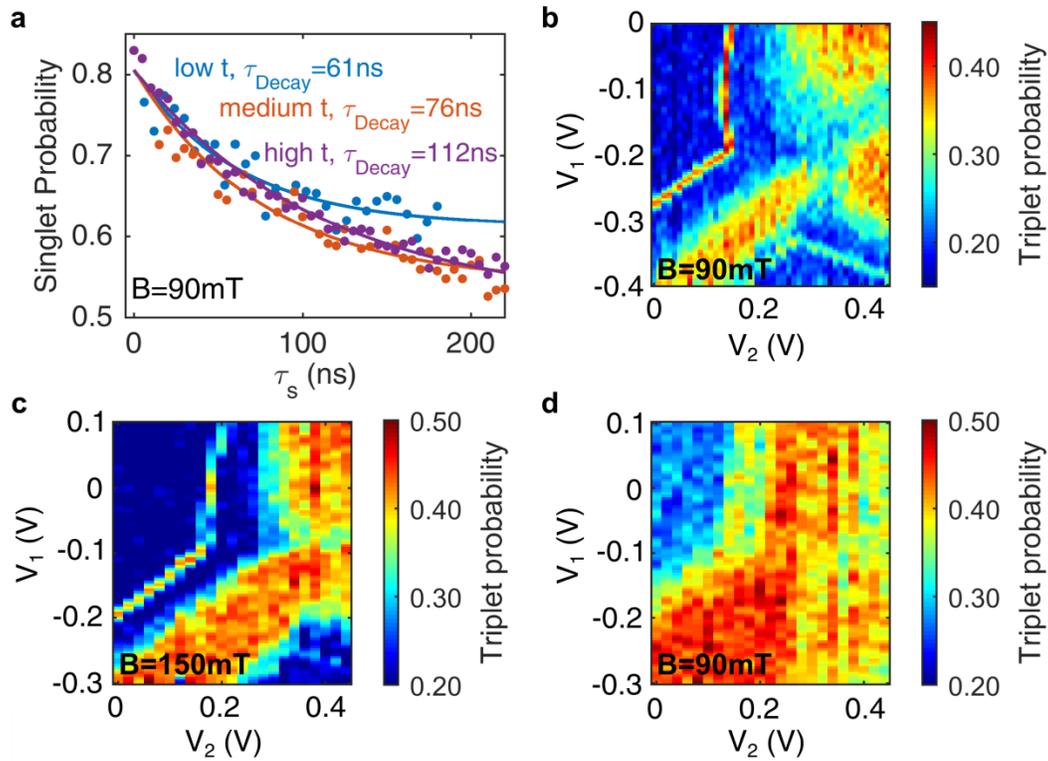

**Supplementary Fig. S5 Influence of the tunnel-coupling on the spin dynamics in the triple quantum dot. a,** Singlet probability as a function of the time $\tau_S$ for the case where the electrons are rotating between separated charge configurations corresponding to the spin map presented in b (violet), c (orange) and d (blue). The data are fitted with an exponential decay with a characteristic time $\tau_{Decay}$. **b, c and d,** Spin mixing maps for decreasing tunnel-coupling between dots. Singlet probability after a 50-ns pulse on $V_1$ and $V_2$.

Even though we observe clear differences in the spin mixing map as discussed in the previous paragraph, no significant change in the increase of the spin coherence time is observed when displacing the electrons (see Fig. S5a). We attribute the slight difference in the measured spin coherence times to a change in the path of the electron during its displacement due to the altered gate voltage configuration. Such a dependence is expected to change the extracted constant $N_{Decay}$ in Fig. 3a. It is worth noting that the displacement data presented in

Fig. 3 (Fig. 2 and 4) correspond to the gate configuration of Fig. S5b (S5c). We therefore conclude that the tunnelling strength is not strongly affecting the spin coherence as long as it is sufficient to allow transfer between the dots.

## VI. Influence of the displacement geometry on the spin coherence time

In the manuscript, the increase of spin coherence is interpreted as a consequence of the electron displacement in a moving quantum dot before and after the tunnelling process. The demonstrated mechanism does therefore not require a particular geometry of the displacement to result in an increase of the spin coherence time. In this section, we provide additional measurements demonstrating that indeed a similar increase for a different transport geometry is observed.

First, we have checked that the rotation direction of the displacement in the triple-dot system was not relevant. In the manuscript, the electrons were rotated anticlockwise along the circular path depicted in Fig. 2. No change of the spin coherence time was observed by rotating clockwise or by alternating clockwise and anticlockwise rotations (data not shown).

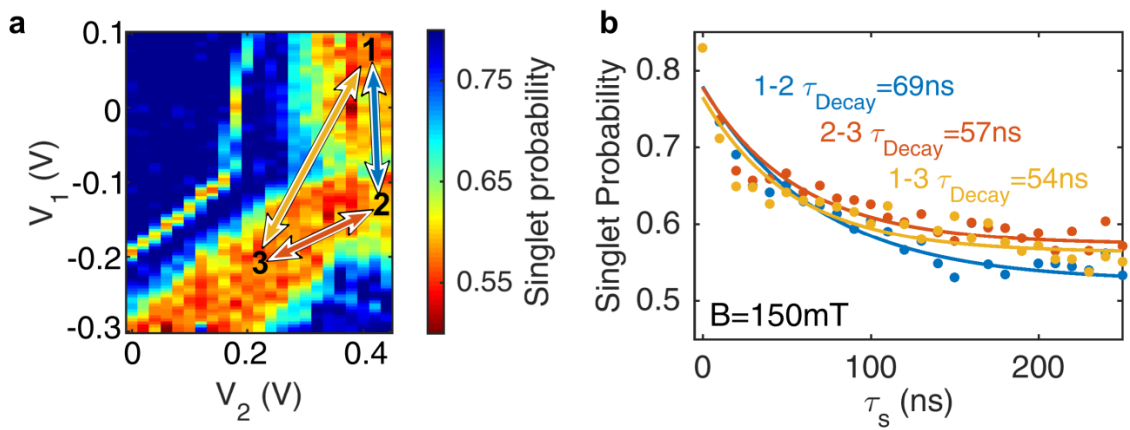

**Supplementary Fig. S6 Influence of the displacement geometry on the spin coherence time. a,** Reproduction of the spin mixing map presented in Fig. S5c with colored arrows indicating the three possible single electron transfers. The system is repeatedly pulsed in between two of the three possible charge configurations to induce arbitrary long displacements of single electrons between two dots. **b,** Singlet probability as a function of the time $\tau_S$ for the case where one electron is displaced between two dots. The time $\tau_r$ spent in each charge configuration during the electron displacement is 2.5 ns, and a magnetic field of 150 mT is applied. A colour code correspondence between the arrows in a and the data in b is used. The data are fitted with an exponential decay with a characteristic time $\tau_{Decay}$.

Second, we have analysed the situation where only one of the two electrons was displaced between two dots. In comparison with the procedure presented in Fig. 2c, we only move one of the electrons back and forth between two static dot configurations after separating the two electrons. The resulting singlet probabilities as a function of the time $\tau_S$ where the electrons are separated are presented in Fig. S6b. They show similar spin dynamics to the case where both electrons are displaced in a circular geometry. Such an observation rules out a scenario where the influence of the nuclei has been cancelled due to electron displacement. Indeed, in a circular displacement, both electrons are experiencing exactly the same effective magnetic fields and as a consequence S and $T_0$ are not expected to mix anymore[20].